\documentclass[aps,prd,reprint,superscriptaddress,nofootinbib,longbibliography]{revtex4-2}

\usepackage{amsmath,amssymb,bm}
\usepackage{booktabs}
\usepackage{graphicx}
\graphicspath{{./}{discrete_bounce_relaxation_L26_h0.8/}{relaxed_bounce_slq_L26_h0.8/}{scaled_potential_continuum_renormalized/}{l36_figures/}}
\usepackage{xcolor}
\usepackage{pgfplots}
\usepackage{placeins}
\usepackage{hyperref}
\pgfplotsset{compat=1.18}
\usetikzlibrary{arrows.meta,positioning}

\newcommand{\dd}{\mathrm{d}}
\newcommand{\Tr}{\operatorname{Tr}}

\begin{document}

\preprint{Working manuscript draft, October 20, 2026}
\title{Higgs Fluctuation Solver in Four Dimensions: Part I---A Matrix-Free
Global-Scalar Bounce and Determinant-Renormalization Audit}

\author{M. S. Cook}
\email{merrittscook1955@gmail.com}
\homepage[]{https://merrittscook1955-ctrl.github.io/in-vero-vacuo/}
\affiliation{In Vero Vacuo LLC}
\date{October 20, 2026}

\begin{abstract}
We report a matrix-free numerical pipeline for semiclassical false-vacuum
decay in a four-component global-$O(4)$ scalar theory in four Euclidean
dimensions. The calculation combines continuum bounce shooting, relaxation
of the sampled profile to a stationary point of the discrete action,
channel-resolved fluctuation spectra, deflated stochastic Lanczos quadrature
(SLQ), and a continuum partial-wave determinant with Gel'fand--Yaglom radial
ratios and large-angular-momentum subtraction. For the benchmark potential,
the continuum bounce action is $B=145.203947$ and the boundary-tail residual
is $0.281955\%$. On an $L=26$, $h=0.8$ lattice, the relaxed field has an RMS
equation-of-motion residual of $4.43\times10^{-12}$; its low spectrum contains
one negative radial mode, four lattice-lifted translations, and three
near-zero internal-orientation modes. The 16-probe bare finite-lattice SLQ
trace--log is $-2079.9042\pm6.4504$, a regulator-dependent quantity. Under
the stated continuum subtraction prescription, the scalar partial-wave
calculation returns a primed trace--log of $40.9324$ at
$\mu=m_F=\sqrt{0.4}$, corresponding to a determinant factor
$1.2931\times10^{-9}$. We show that the leading $\rho^8$ potential is not
closed under its scalar one-loop divergence: local $\rho^{10}$ and
$\rho^{12}$ counterterms are required, and their finite renormalized
coefficients have not been matched. Thus the reported determinant is a
prescription-level result, not a complete matched EFT prediction. Its quoted
$0.0300$ uncertainty is an estimated hard-coded value rather than an
independently established error bar. The
translation-normalized scalar semiclassical expression is
$5.9974\times10^{-70}$ in lattice energy units to the fourth power; it omits
the internal-orientation Jacobian and is not a physical electroweak or gauged
decay rate. We summarize tested $SU(2)$ operator, vertex, and pole-algebra
infrastructure, while identifying finite full-gauge matching, gauge-sector
determinant completion, and controlled continuum/volume limits as open work.
\end{abstract}

\maketitle

\section{Introduction}
\label{sec:introduction}

Semiclassical false-vacuum decay is governed by a nontrivial Euclidean saddle
and by the fluctuation determinant around that saddle. The saddle action
controls the leading exponential, whereas the one-loop prefactor requires a
careful treatment of the negative mode, translation collective coordinates,
internal zero modes, and ultraviolet renormalization
\cite{Coleman1977,CallanColeman1977}. These ingredients are individually
well understood in analytic and partial-wave calculations, but their
simultaneous numerical treatment becomes demanding when the fluctuation
operator is large, the background is spatially varying, and lattice and
continuum regulators must be related.

This work documents a computational pipeline developed for a four-component
real scalar field with a global $O(4)$ symmetry. The production result is a
scalar-theory benchmark, not a calculation of the Standard Model Higgs decay
rate. The solver combines an $O(4)$-symmetric continuum bounce with a
four-dimensional periodic lattice representation, matrix-free Hessian
operators, channel-by-channel low-mode extraction, a stochastic trace--log,
and a continuum-subtracted radial determinant. The purpose of Part I is to
describe this scalar benchmark and the verified computational components
without conflating finite-lattice diagnostics with a physical decay
prefactor.

Recent work has advanced several neighboring parts of Higgs-vacuum decay.
Perturbative studies have treated the Standard Model bounce with running
couplings and gravity, computed complete one-loop rates in scalar
quartic--quartic models, analyzed gauge--Goldstone mixing and gradient effects
in gauged scalar theories, developed non-thin-wall one-loop corrections for
scalar bounces, and derived four-dimensional determinant-renormalization
formulas for fields of spin 0, $1/2$, and 1
\cite{Espinosa2020,GuadaNemevsek2020,AiCruzGarbrechtTamarit2020,
AiAlexandreSarkar2024,BaratellaEtAl2025}.
A recent Standard Model calculation has also revisited the full one-loop
gauge prefactor and transverse-mode counting \cite{BaratellaGauge2025}.
Lattice-based Higgs-vacuum studies have examined real-time stability during
post-inflationary kination; a distinct Euclidean-lattice Monte Carlo line has
proposed decay estimators demonstrated in metastable quantum-mechanical
systems \cite{LaverdaRubio2024,ShenDraperElKhadra2023,JinSwaim2025}. The
present calculation is
complementary to these efforts: it combines a continuum $O(4)$ bounce and
continuum-subtracted radial determinant with a relaxed finite-lattice
background, matrix-free Hessian analysis, and a bare SLQ trace--log diagnostic.
It does not include the Standard Model running-potential and gravitational
effects, constitute a Euclidean field-theory Monte Carlo decay measurement,
or provide a fully matched gauge-sector determinant. The scalar result is a
scaled global-$O(4)$ benchmark, with the finite $\rho^{10}$ and $\rho^{12}$
coefficients still unmatched, rather than a precision electroweak decay rate.

The project also contains a growing quadratic and perturbative $SU(2)$
extension. Several operator identities, homogeneous limits, vertex
generators, and ultraviolet pole relations have been tested. The scalar
one-loop analysis additionally shows that the tree-level potential truncated
at $\rho^8$ is not closed under the retained fluctuation determinant:
$\rho^{10}$ and $\rho^{12}$ counterterms are required, and their finite
coefficients have not been matched. Selected finite-matching pilots have
been evaluated, but the finite gauge-sector coefficient set and a complete
gauge--Goldstone--ghost determinant are not available. We therefore treat
the determinant and matching results as implementation status and
prerequisite checks, not as a fully matched EFT or gauged prediction. A later
manuscript(Part II) would be justified only after those missing calculations and
their systematic checks are completed.

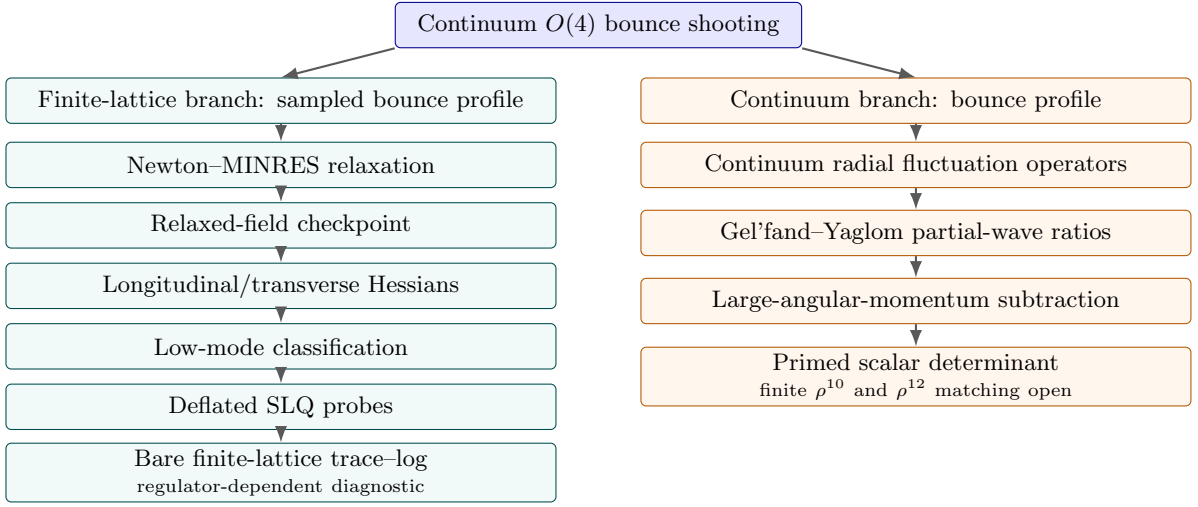
\begin{figure*}[t]
\centering
\begin{tikzpicture}[
  >=Latex,
  flow/.style={draw=blue!55!black, fill=blue!4, rounded corners=2pt,
    text width=7.1cm, minimum height=6mm, inner sep=2.5pt, align=center,
    font=\small},
  root/.style={flow, text width=5.2cm, fill=blue!10},
  lattice/.style={flow, draw=teal!60!black, fill=teal!5},
  continuum/.style={flow, draw=orange!70!black, fill=orange!7},
  arr/.style={-{Latex[length=2.2mm]}, thick, draw=black!65}
]
\node[root] (shoot) at (0,0) {Continuum $O(4)$ bounce shooting};

\node[lattice] (sample) at (-4.2,-1.0)
  {Finite-lattice branch: sampled bounce profile};
\node[continuum] (profile) at (4.2,-1.0)
  {Continuum branch: bounce profile};
\draw[arr] (shoot.south west) -- (sample.north);
\draw[arr] (shoot.south east) -- (profile.north);

\node[lattice] (relax) at (-4.2,-1.85)
  {Newton--MINRES relaxation};
\node[lattice] (checkpoint) at (-4.2,-2.65)
  {Relaxed-field checkpoint};
\node[lattice] (hessian) at (-4.2,-3.45)
  {Longitudinal/transverse Hessians};
\node[lattice] (lowmodes) at (-4.2,-4.25)
  {Low-mode classification};
\node[lattice] (slq) at (-4.2,-5.05)
  {Deflated SLQ probes};
\node[lattice] (bare) at (-4.2,-5.92)
  {Bare finite-lattice trace--log\\[-1pt]
   \scriptsize regulator-dependent diagnostic};
\draw[arr] (sample) -- (relax);
\draw[arr] (relax) -- (checkpoint);
\draw[arr] (checkpoint) -- (hessian);
\draw[arr] (hessian) -- (lowmodes);
\draw[arr] (lowmodes) -- (slq);
\draw[arr] (slq) -- (bare);

\node[continuum] (radial) at (4.2,-1.85)
  {Continuum radial fluctuation operators};
\node[continuum] (gy) at (4.2,-2.75)
  {Gel'fand--Yaglom partial-wave ratios};
\node[continuum] (subtraction) at (4.2,-3.65)
  {Large-angular-momentum subtraction};
\node[continuum] (determinant) at (4.2,-4.65)
  {Primed scalar determinant\\[-1pt]
   \scriptsize finite $\rho^{10}$ and $\rho^{12}$ matching open};
\draw[arr] (profile) -- (radial);
\draw[arr] (radial) -- (gy);
\draw[arr] (gy) -- (subtraction);
\draw[arr] (subtraction) -- (determinant);
\end{tikzpicture}
\caption{Overview of the scalar calculation. The continuum bounce seeds the
finite-lattice relaxation and is also the input to the continuum radial
calculation. The finite-lattice SLQ trace--log is a regulator-dependent
diagnostic; the partial-wave route gives the quoted continuum-subtracted
determinant under the stated prescription. These are complementary routes,
not successive steps, and the determinant is not a complete matched EFT
prediction because the finite $\rho^{10}$ and $\rho^{12}$ counterterms remain
unmatched.}
\label{fig:data-flow}
\end{figure*}

The calculation's main stages and the distinct roles of its lattice and
continuum determinant analyses are summarized in Fig.~\ref{fig:data-flow}.
The two determinant routes are complementary rather than sequential: the
bare lattice trace--log is a regulator-dependent diagnostic, whereas the
continuum partial-wave calculation supplies the reported subtraction-
prescription result.

\section{Model and semiclassical observable}
\label{sec:model}

\begin{figure*}[t]
\centering
\includegraphics[width=0.92\textwidth]{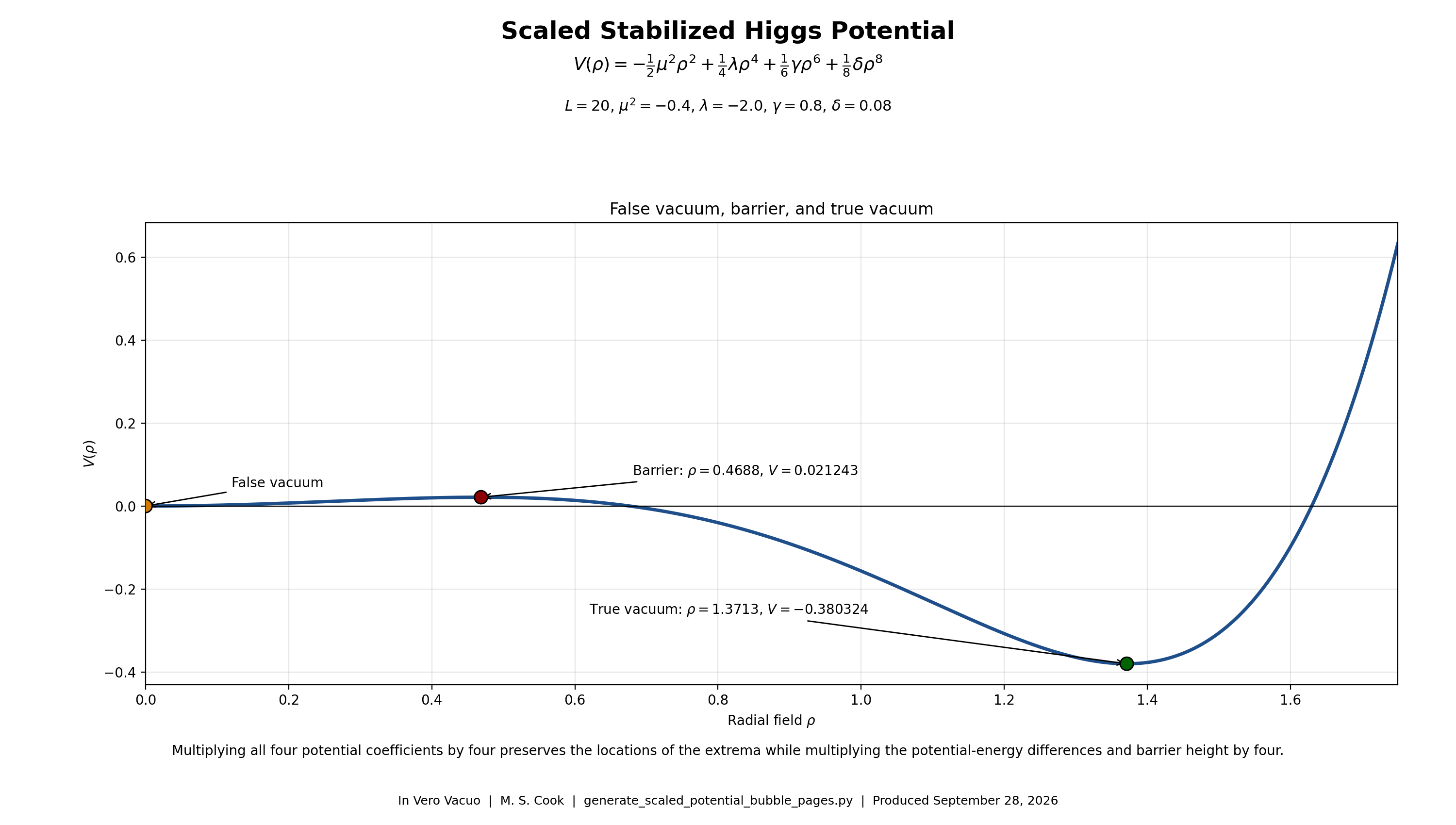}
\caption{Benchmark radial potential in the solver's scaled units. The plot
marks the false vacuum at $\rho=0$, the barrier, and the deeper nonzero
minimum for the coefficients in Eq.~\eqref{eq:parameters}. This is the
potential-shape benchmark used to define the continuum bounce problem.}
\label{fig:benchmark-potential}
\end{figure*}

We consider a real four-component field $\bm\Phi$ with radial amplitude
$\rho=(\bm\Phi\cdot\bm\Phi)^{1/2}$ and Euclidean action
\begin{equation}
 S_E[\bm\Phi]=\int\dd^4x\left[
 \frac{1}{2}\partial_\mu\bm\Phi\cdot\partial_\mu\bm\Phi
 +V(\rho)\right].
 \label{eq:action}
\end{equation}
The benchmark potential is
\begin{equation}
 V(\rho)=-\frac{\mu^2}{2}\rho^2+
 \frac{\lambda}{4}\rho^4+
 \frac{\gamma}{6}\rho^6+
 \frac{\delta}{8}\rho^8,
 \label{eq:potential}
\end{equation}
with
\begin{equation}
 \mu^2=-0.40,\qquad \lambda=-2.00,\qquad
 \gamma=0.80,\qquad \delta=0.08.
 \label{eq:parameters}
\end{equation}
The resulting false-vacuum minimum, barrier, and deeper minimum are shown in
Fig.~\ref{fig:benchmark-potential}.
The false vacuum is at $\rho=0$, with $m_F^2=V''(0)=-\mu^2=0.4$.
The negative quartic produces a barrier, while the higher powers stabilize
the potential at large field amplitude. All benchmark values are quoted in
the solver's scaled units.

Under the usual assumptions for scalar false-vacuum decay, the least-action
solution is $O(4)$ symmetric \cite{ColemanGlaserMartin1978} and obeys
\begin{equation}
\rho_B''(r)+\frac{3}{r}\rho_B'(r)=V'(\rho_B),\\
\rho_B'(0)=0,\qquad \rho_B(\infty)=0.
 \label{eq:bounce}
\end{equation}
Its action relative to the false vacuum is
\begin{equation}
 B=2\pi^2\int_0^\infty r^3\left[
 \frac{1}{2}\rho_B'(r)^2+V(\rho_B(r))-V(0)\right]\dd r.
 \label{eq:bounce-action}
\end{equation}

For a scalar theory, the semiclassical expression has the schematic form
\begin{equation}
 \frac{\Gamma}{V_4}=
 \left(\frac{B}{2\pi}\right)^2
 \left|\frac{\det{}'H_B}{\det H_F}\right|^{-1/2}e^{-B}
 \,\mathcal J_{\mathrm{internal}}\,[1+O(\hbar)],
 \label{eq:rate}
\end{equation}
where $H_B$ and $H_F$ are the bounce and false-vacuum Hessians. The prime
denotes removal of collective-coordinate eigenvalues, and the unique negative
mode is retained through the absolute determinant. Equation~\eqref{eq:rate}
is included to define the target observable; the calculation reported here
does not determine $\mathcal J_{\mathrm{internal}}$ or a physical energy
conversion for the rate.

\section{Numerical methods}
\label{sec:methods}

\subsection{Continuum bounce}
\label{subsec:continuum-bounce}

The bounce equation is integrated from a small positive $\epsilon$ using the
regular origin expansion
\begin{equation}
 \rho_B(\epsilon)=\rho_0+\frac{V'(\rho_0)}{8}\epsilon^2,
 \qquad
 \rho_B'(\epsilon)=\frac{V'(\rho_0)}{4}\epsilon.
\end{equation}
An overshoot/undershoot bisection in $\rho_0$ isolates the separatrix
approaching the false vacuum. Integration terminates when the field enters a
specified physical tail; this avoids extending an approximate separatrix
beyond the range where numerical roundoff remains controlled. The action is
then evaluated from Eq.~\eqref{eq:bounce-action} using the $S^3$ angular
measure $2\pi^2r^3$.

\subsection{Periodic lattice and matrix-free Hessian}
\label{subsec:lattice-hessian}

The lattice uses periodic boundary conditions and a nearest-neighbor
four-dimensional Laplacian,
\begin{equation}
 (-\Delta_h f)_x=\frac{1}{h^2}\left[
 8f_x-\sum_{\mu=1}^{4}
 (f_{x+\hat\mu}+f_{x-\hat\mu})\right].
 \label{eq:lattice-laplacian}
\end{equation}
The stencil is applied directly to reshaped arrays; the global Hessian is
exposed as a matrix-free linear operator rather than assembled. At $L=26$,
one scalar channel has $L^4=456{,}976$ sites and the four-component field has
$1{,}827{,}904$ degrees of freedom. The sampled continuum profile is relaxed
against the discrete action using a Newton--MINRES method. The false-vacuum
operator is diagonalized analytically in the periodic Fourier basis, which
also provides a reference for checking the lattice indexing and degeneracies.

For a radial background, the Hessian splits into one longitudinal and three
degenerate transverse channels,
\begin{align}
 H_L&=-\Delta_h+U_L(x), &
 U_L&=-\mu^2+3\lambda\rho^2+5\gamma\rho^4+7\delta\rho^6,\\
 H_T&=-\Delta_h+U_T(x), &
 U_T&=-\mu^2+\lambda\rho^2+\gamma\rho^4+\delta\rho^6.
 \label{eq:hessian-channels}
\end{align}
The transverse degeneracy is applied analytically, while eigenpairs are
computed separately in each channel. This avoids cutting through a degenerate
multiplet during low-mode extraction.

\subsection{Lattice-to-physical scaling and run modes}
\label{subsec:lattice-physical-scaling}

The solver has two distinct spacing conventions. Exploratory research mode
uses an explicit, dimensionful lattice spacing $a$, with conversions
\begin{equation}
\begin{aligned}
 \Phi_{\mathrm{phys}}&=\frac{\Phi_{\mathrm{lat}}}{a},\\
 m_{\mathrm{phys}}^2&=\frac{m_{\mathrm{lat}}^2}{a^2},\\
 V_{4,\mathrm{phys}}&=(La)^4,\\
 \Lambda_{\mathrm{UV}}&\sim\frac{\pi}{a}.
\end{aligned}
\label{eq:research-mode-scaling}
\end{equation}
By contrast, the scalar production results reported here---including bounce
relaxation, Hessian spectra, and SLQ---use dimensionless solver units and a
refinement spacing $h$; the main benchmark has $h=0.8$. The spacing campaign
varies $h$ while holding the box extent $Lh$ fixed in these scaled units.
Thus $h$ is not the dimensionful spacing $a$ in
Eq.~\eqref{eq:research-mode-scaling}, and no substitution $a=h$ is implied.
The continuum partial-wave calculation uses $m_F$ as its reference scale in
the solver's scaled convention. Consequently, the scalar prefactor remains
in lattice-energy units to the fourth power; conversion to a physical rate in
$\mathrm{GeV}^4$ requires an explicit physical scale and is not part of the
reported production calculation.

\subsection{One-loop closure of the radial potential}
\label{subsec:potential-closure}

The tree-level benchmark in Eq.~\eqref{eq:potential} is truncated at
$\rho^8$, but that truncation is not closed under the scalar one-loop
divergence when the $\rho^6$ and $\rho^8$ interactions are retained
nonperturbatively in the leading fluctuation operators. For the radial
channels in Eq.~\eqref{eq:hessian-channels}, the local scalar divergence is
proportional to $U_L^2+3U_T^2$. Writing $m_F^2=-\mu^2$, its field-dependent
polynomial is
\begin{align}
U_L^2+3U_T^2={}&4m_F^4+12\lambda m_F^2\rho^2
+(12\lambda^2+16\gamma m_F^2)\rho^4\notag\\
&+(36\lambda\gamma+20\delta m_F^2)\rho^6
+(28\gamma^2+48\lambda\delta)\rho^8\notag\\
&+76\gamma\delta\rho^{10}+52\delta^2\rho^{12}.
\label{eq:radial-pole-polynomial}
\end{align}
The field-independent term cancels in the bounce-minus-false-vacuum action.
The final two terms do not belong to the original potential, so a
renormalization-closed scalar basis at this loop order must also allow
\begin{equation}
V_{\mathrm{EFT}}(\rho;\mu_R)=V(\rho;\mu_R)
+\frac{c_{10}(\mu_R)}{10}\rho^{10}
+\frac{c_{12}(\mu_R)}{12}\rho^{12}.
\label{eq:extended-radial-potential}
\end{equation}
This is an order-by-order closure statement: $c_{10}$ and $c_{12}$ are
one-loop-order counterterm coefficients and, in the adopted counting, do not
enter the leading bounce equation or Hessian. Their finite renormalized
values nevertheless require matching conditions and have not been fixed in
the benchmark reported here. Promoting these operators to independent
leading interactions would require a still larger basis at higher orders:
for example, the curvature of a leading $\rho^{12}$ term scales as
$\rho^{10}$, whose square generates a $\rho^{20}$ contribution.
This scalar result is not a closure proof for the full gauged theory, which
also has derivative and gauge-sector counterterms. The need to include
loop-generated local operators in the EFT basis is standard in
order-by-order effective-theory renormalization
\cite{Vassilevich2003,BanerjeeChakraborttyRamkumar2024}.

\subsection{Low modes and stochastic trace--log}
\label{subsec:slq}

For a positive operator, $\log\det H=\Tr\log H$. On the bounce, the
negative and collective-coordinate modes require explicit classification;
the low eigenpairs are therefore computed and deflated before estimating the
remaining trace. For independent Rademacher probes $z_k$, the trace is
estimated by
\begin{equation}
 \Tr f(H)\simeq\frac{1}{N_p}\sum_{k=1}^{N_p}z_k^T f(H)z_k,
 \qquad f(H)=\log H.
 \label{eq:trace-estimator}
\end{equation}
The probe average is a Hutchinson-type stochastic trace estimator;
Lanczos iteration maps each quadratic form to a Gaussian quadrature on a
small tridiagonal matrix \cite{Hutchinson1989,UbaruChenSaad2017}. The same
probe stream is used for the bounce and false-vacuum operators so that their
difference benefits from correlated stochastic fluctuations. Exact low
modes are restored analytically after projection.

\subsection{Continuum partial-wave determinant}
\label{subsec:continuum-det}

For an $O(4)$ background, the radial operators are organized by angular
momentum $\ell$ with degeneracy $g_\ell=(\ell+1)^2$. Radial Gel'fand--Yaglom
ratios replace functional determinants by ratios of solutions of initial
value problems \cite{GelfandYaglom1960,Dunne2008}; the large-$\ell$ asymptotics are
subtracted before summation \cite{DunneKirsten2006}. Translation and
orientation zero modes are handled by a positive spectral shift, division by
the known powers of that shift, and extrapolation toward zero, following
standard primed-determinant treatments \cite{McKaneTarlie1995,
KirstenMcKane2004,BaackeLavrelashvili2004}.

\section{Scalar benchmark results}
\label{sec:scalar-results}

\subsection{Bounce and discrete stationary field}
\label{subsec:bounce-results}

\begin{figure*}[t]
\centering
\includegraphics[width=0.92\textwidth]{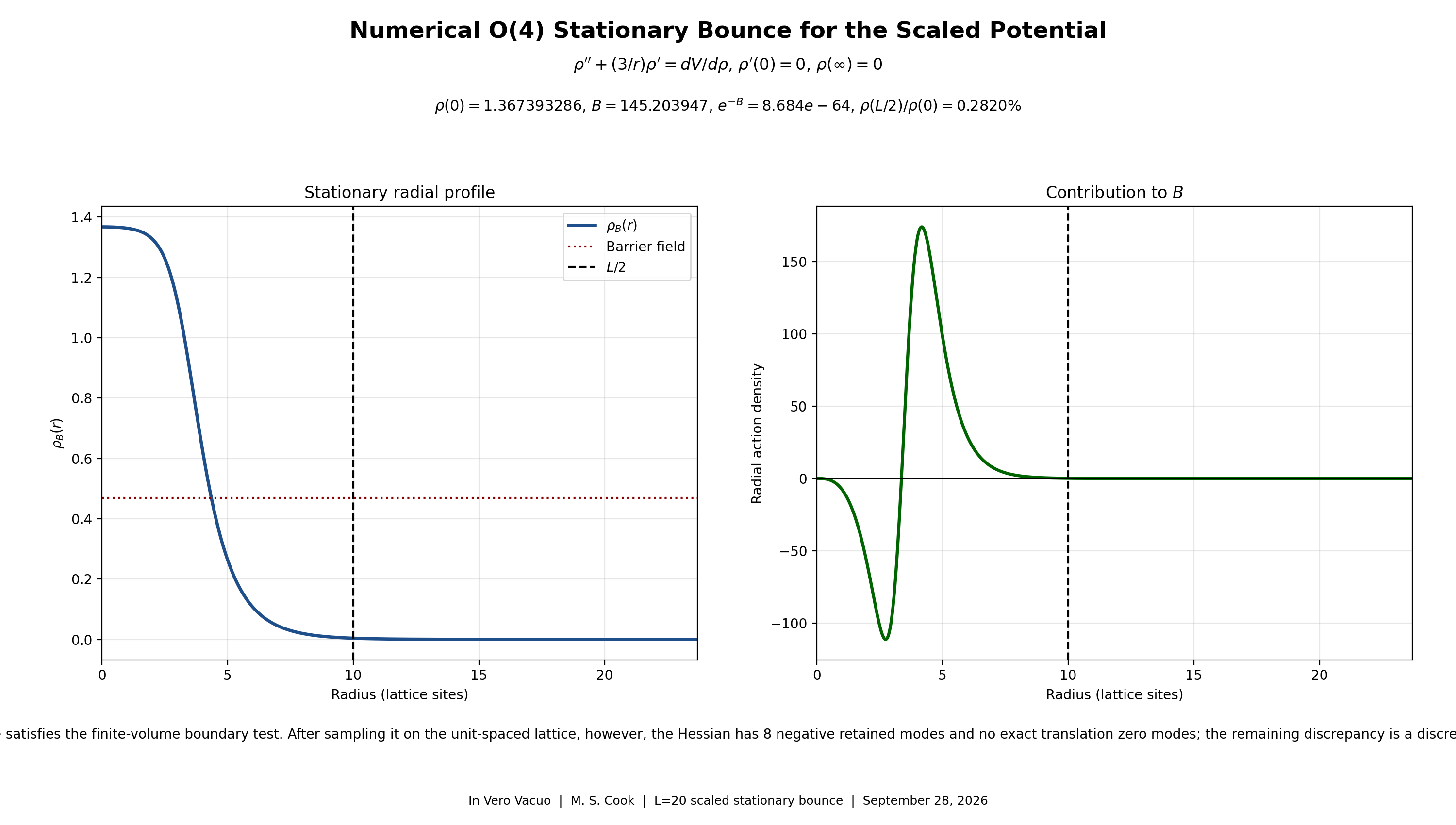}
\caption{Continuum $O(4)$ bounce profile and radial contribution to the
action for the benchmark potential. The profile begins at
$\rho_B(0)=1.367393286$ and approaches the false vacuum in the tail; the
dashed vertical line marks the half-box radius used in the $L=20$ boundary
test. The reported boundary-tail residual is $0.2820\%$.}
\label{fig:continuum-bounce}
\end{figure*}

For Eqs.~\eqref{eq:potential}--\eqref{eq:parameters}, the continuum shooting
solution has center value $\rho_B(0)=1.367393286$ and action
$B=145.2039474492$. The reported finite-box boundary residual is
$0.281955\%$, below the adopted one-percent criterion. The leading tunneling
factor is $e^{-B}=8.6841\times10^{-64}$. The radial bounce and radial action
density for this continuum solution are shown in Fig.~\ref{fig:continuum-bounce}.

\begin{figure*}[t]
\centering
\includegraphics[width=0.92\textwidth]{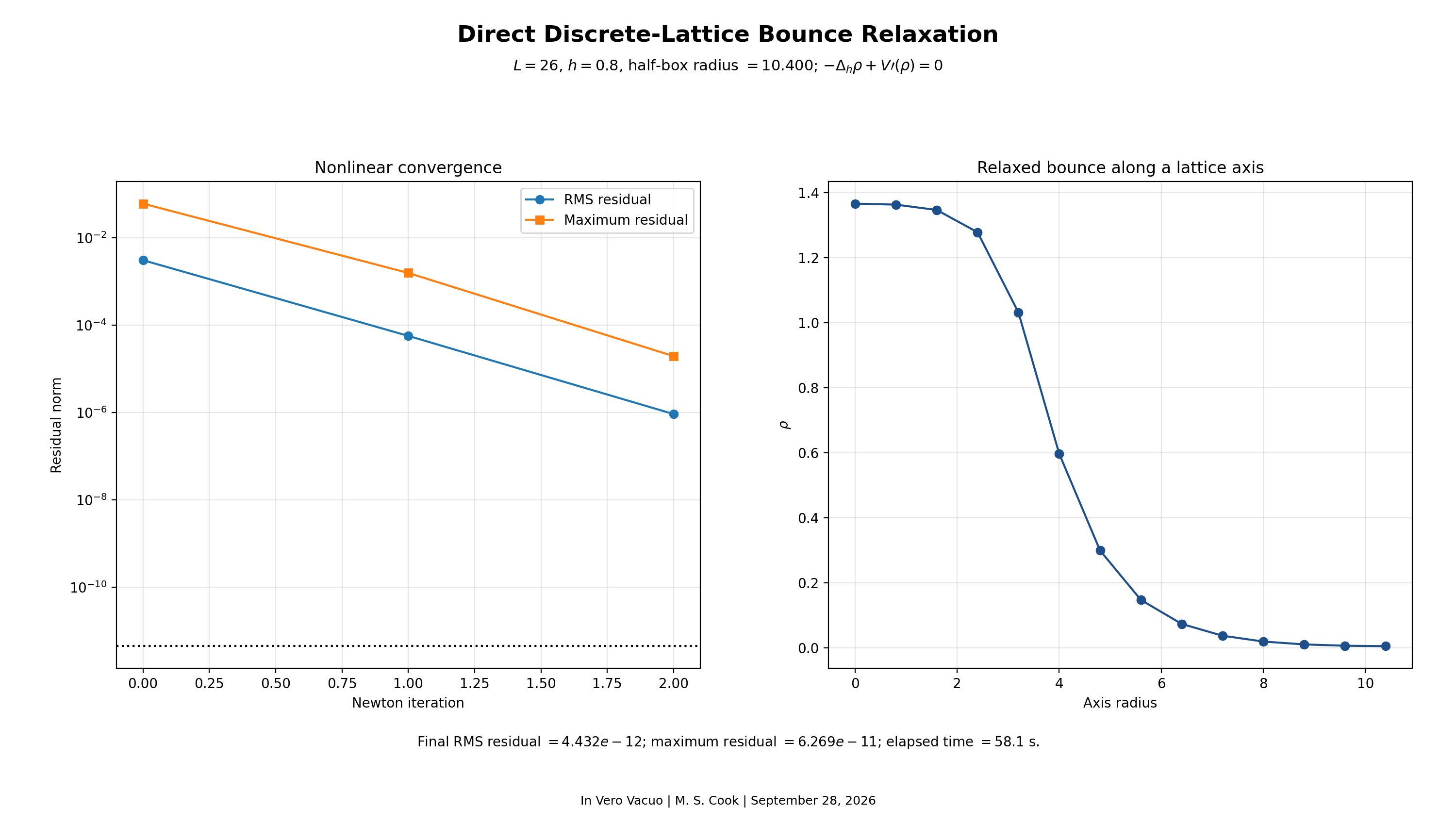}
\caption{Direct relaxation of the sampled bounce on the periodic $L=26$,
$h=0.8$ lattice. The left panel shows the decrease of the RMS and maximum
equation-of-motion residuals during Newton--MINRES iterations; the right
panel shows the converged radial field along a lattice axis. The final RMS
residual is approximately $4.43\times10^{-12}$.}
\label{fig:discrete-relaxation}
\end{figure*}

The lattice calculation uses $L=26$ and refinement spacing $h=0.8$. For the
regenerated relaxed field, the RMS and maximum discrete equation-of-motion
residuals are $4.43266\times10^{-12}$ and $6.30906\times10^{-11}$,
respectively. The identity relating the transverse Hessian to the residual
provides an additional check of the internal-orientation zero mode.
The relaxation history and the relaxed field along a lattice axis are shown
in Fig.~\ref{fig:discrete-relaxation}.

\subsection{Collective-coordinate spectrum}
\label{subsec:spectrum-results}

\begin{figure*}[t]
\centering
\includegraphics[width=0.92\textwidth]{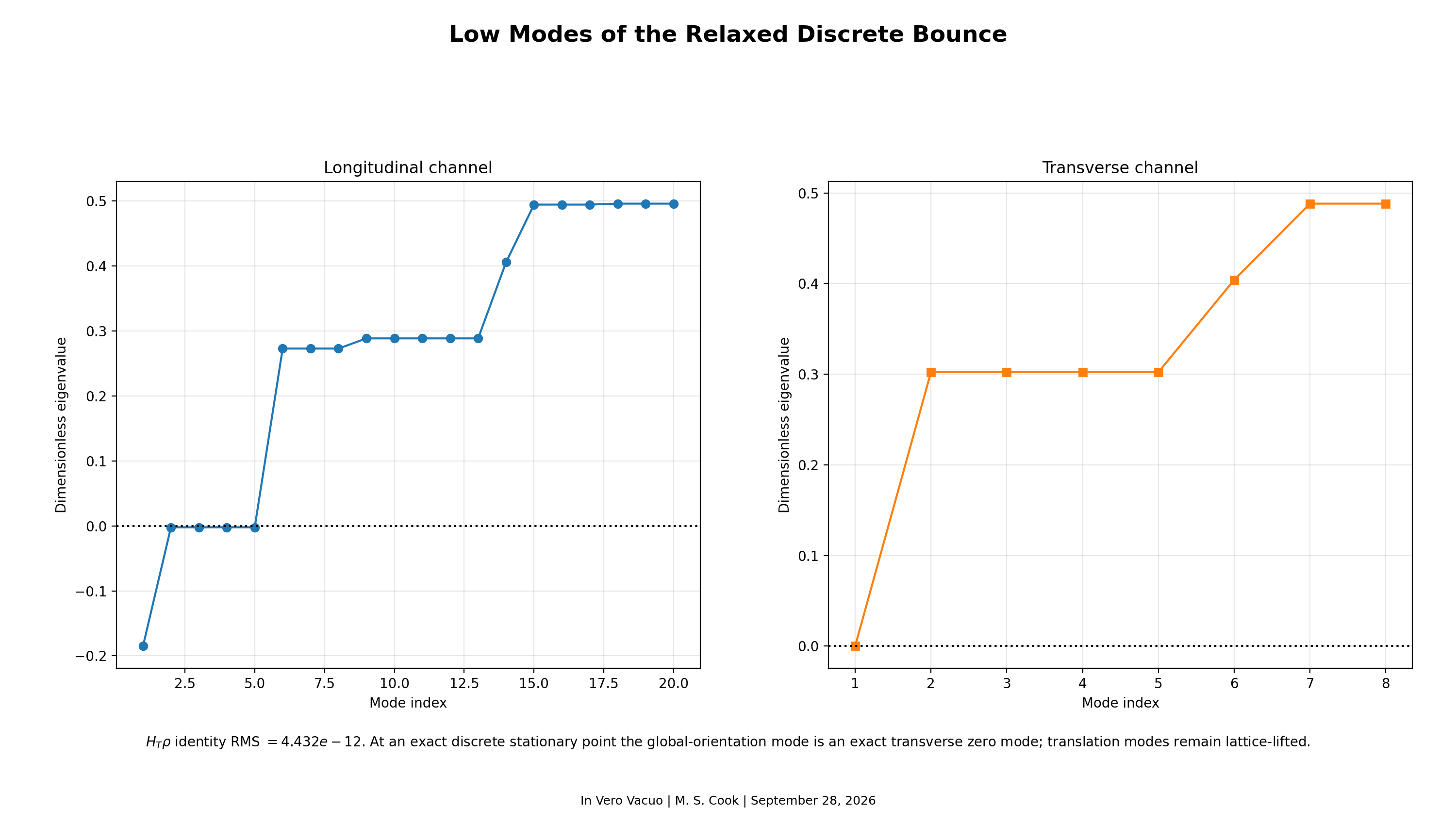}
\caption{Low eigenvalues of the relaxed discrete-bounce Hessian in the
longitudinal and transverse channels. The longitudinal channel contains one
negative mode and four near-zero translation modes; the transverse channel
contains the global-orientation zero mode, with its threefold internal
degeneracy. The dotted lines indicate zero eigenvalue.}
\label{fig:low-modes}
\end{figure*}

The low spectrum has the expected pattern: one negative longitudinal mode,
four translation modes lifted slightly below zero by the finite lattice, and
three internal-orientation modes close to zero. The reported eigenvalues are
summarized in Table~\ref{tab:scalar-results}. The translation eigenvalues
are not exact zeros because continuous translation symmetry is broken by the
fixed lattice. The orientation triplet is protected by the global symmetry
and is numerically zero at the relaxation tolerance. The channel-resolved low
eigenvalues are visualized in Fig.~\ref{fig:low-modes}.

\begin{table}[t]
\caption{Principal scalar benchmark results. Values are in the solver's
scaled units unless stated otherwise. The continuum determinant is a
prescription-level result because the finite $c_{10}$ and $c_{12}$ matching
coefficients are not fixed. Its $0.0300$ uncertainty is the source's estimate,
not an independently calibrated error.}
\label{tab:scalar-results}
\begin{ruledtabular}
\footnotesize
\begin{tabular}{@{}p{0.47\columnwidth}p{0.47\columnwidth}@{}}
Quantity & Result \\
\hline
Continuum center field & $1.367393286$ \\
Continuum bounce action & $145.2039474492$ \\
Boundary residual & $0.281955\%$ \\
Discrete RMS / max residual & $4.43\times10^{-12}$ / $6.31\times10^{-11}$ \\
Negative longitudinal mode & $-0.1850415$ \\
Four translation modes & $-0.00212653$ each \\
Three orientation modes & $-5.33\times10^{-12}$ each \\
Bare finite-lattice SLQ trace--log & $-2079.9042\pm6.4504$ \\
Continuum-subtracted scalar trace--log & $40.9324$ (estimated $\pm0.0300$) \\
Continuum-subtracted determinant factor & $1.2931\times10^{-9}$ \\
\end{tabular}
\normalsize
\end{ruledtabular}
\end{table}

\subsection{Bare SLQ and continuum-subtracted determinant}
\label{subsec:det-results}

\begin{figure}[t]
\centering
\includegraphics[width=\columnwidth]{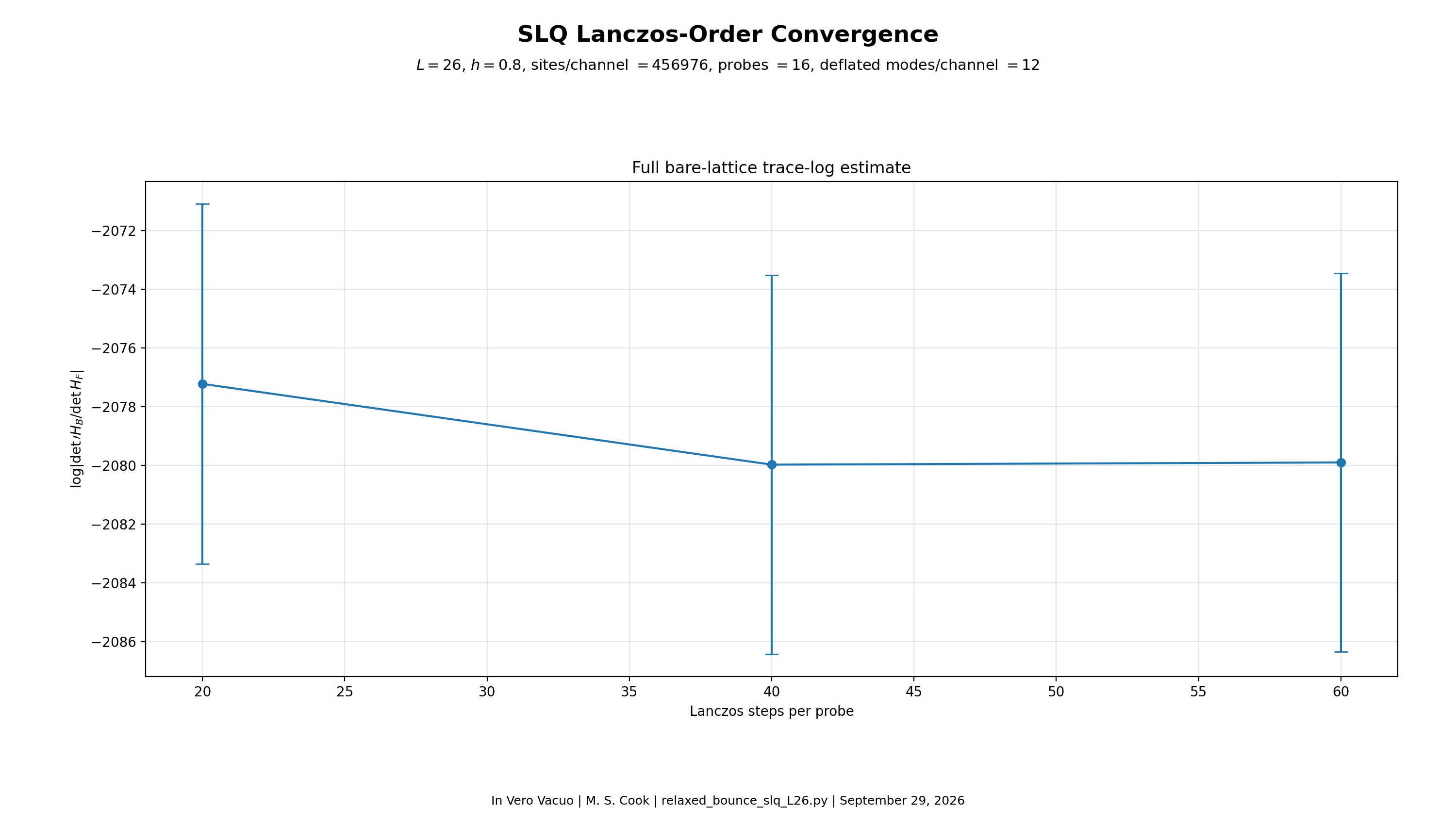}
\caption{Lanczos-order dependence of the bare finite-lattice trace--log
estimate for $L=26$, $h=0.8$, 16 Rademacher probes, and 12 deflated
eigenpairs per channel. Error bars show the reported estimator uncertainty.
The plotted quantity remains regulator dependent and is not the
continuum-renormalized determinant.}
\label{fig:slq-convergence}
\end{figure}

The production SLQ calculation used 16 Rademacher probes, 12 deflated
eigenpairs per bounce channel, and Lanczos orders 20, 40, and 60. The order-60
bare finite-lattice trace--log was
\begin{equation}
 \left.\log\left|\frac{\det{}'H_B}{\det H_F}\right|\right|_{\rm bare}
 =-2079.9042\pm6.4504.
 \label{eq:bare-trace}
\end{equation}
The quoted spread is the reported finite-lattice estimator uncertainty. This
quantity is dominated by ultraviolet regulator dependence and is not a
physical enhancement of the decay rate. It cannot be compared directly with
the continuum-renormalized result without matching the regulator and
counterterms. Figure~\ref{fig:slq-convergence} shows the bare estimate as a
function of Lanczos order; its error bars emphasize the stochastic spread at
the probe count used here.

\begin{figure*}[t]
\centering
\includegraphics[width=0.92\textwidth]{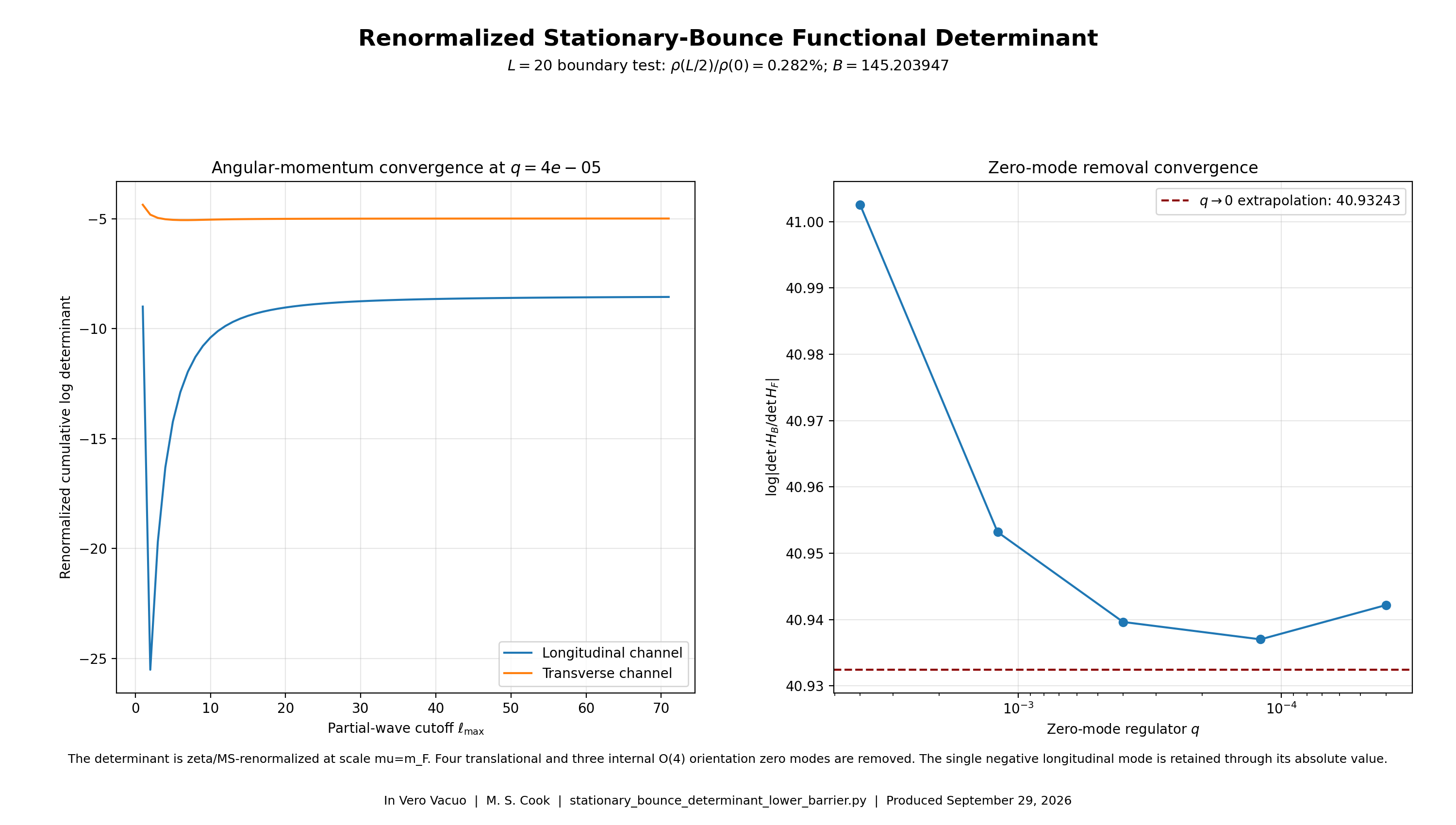}
\caption{Continuum determinant convergence diagnostics for the $L=20$
stationary-bounce benchmark. The left panel shows the asymptotically
subtracted partial-wave sum as the angular-momentum cutoff increases; the
right panel shows the positive-regulator extrapolation used to remove the
four translation and three internal-orientation zero modes. The plotted
$q\to0$ intercept is approximately $40.9324$. The source's quoted
$0.0300$ uncertainty is an estimated, hard-coded value, not an uncertainty
inferred from this figure.}
\label{fig:continuum-determinant}
\end{figure*}

The separate continuum calculation uses a radial cutoff of 20,
$\ell_{\max}=70$, and renormalization scale $\mu=m_F=\sqrt{0.4}$. It reports
\begin{align}
 \log\left|\frac{\det{}'H_B}{\det H_F}\right|_{\overline{\rm MS}}
 &=40.9324,\\
 \left|\frac{\det{}'H_B}{\det H_F}\right|^{-1/2}_{\overline{\rm MS}}
 &=1.2931\times10^{-9}.
 \label{eq:renormalized-det}
\end{align}
The source records an uncertainty estimate of $0.0300$ for the trace--log,
but that value is hard-coded and is not the output of a complete systematic
error analysis. Accordingly, the central value is a benchmark result of the
implemented prescription; its precision should not be overstated.
There is also a separate renormalization-closure limitation: with nonzero
$\gamma$ and $\delta$, Eq.~\eqref{eq:radial-pole-polynomial} contains local
$\rho^{10}$ and $\rho^{12}$ terms. The determinant's asymptotic subtraction
produces a finite number for the specified operator, but the finite
renormalized coefficients $c_{10}(\mu_R)$ and $c_{12}(\mu_R)$ have not been
fixed by matching. Thus $40.9324$ is a prescription-level determinant
result for the chosen leading operator, not a complete matched one-loop EFT
prediction.
The partial-wave sum and positive-shift zero-mode extrapolation are shown in
Fig.~\ref{fig:continuum-determinant}.

Including the continuum bounce action and the four-translation factor gives
the scalar partial expression
\begin{align}
 \left(\frac{B}{2\pi}\right)^2
 \left|\frac{\det{}'H_B}{\det H_F}\right|^{-1/2}_{\overline{\rm MS}}
 e^{-B}&=5.9974\times10^{-70},\\
 &\qquad\text{(lattice energy units)}^4.
 \label{eq:scalar-partial-prefactor}
\end{align}
Equation~\eqref{eq:scalar-partial-prefactor} omits the internal-orientation
group-volume Jacobian and does not include an explicit conversion to a
physical energy scale. It is therefore not a complete scalar decay rate, and
it is not a gauged-Higgs prediction. It also inherits the unresolved finite
$c_{10}$ and $c_{12}$ matching in the one-loop scalar potential.

\section{Status of the quadratic and perturbative \texorpdfstring{$SU(2)$}{SU(2)} extension}
\label{sec:gauge-status}

The solver includes a research branch for quadratic gauge--Goldstone and
ghost operators around a radial Higgs background with zero background gauge
field. The gauge operator is derived from lattice links rather than from a
centered continuum derivative; positive background-$R_\xi$ gauges are
supported. Homogeneous Fourier spectra, normalization conventions, and
Ward-identity residuals have been tested at $\xi=0.5,1,2$. In a regenerated
$L=26$ stationary background, the largest reported on-shell Ward residual
was $5.52\times10^{-14}$. The homogeneous regulated determinant test also
approaches gauge-parameter independence as its auxiliary IR shift is
removed.

\subsection{Wilson plaquette action and vertex convention}
\label{subsec:wilson-plaquette}

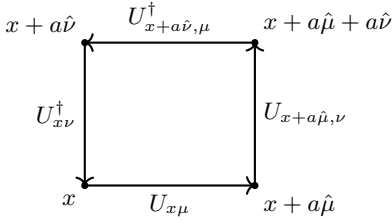
\begin{figure}[t]
\centering
\begin{tikzpicture}[scale=0.9,every node/.style={font=\small}]
  \coordinate (x) at (0,0);
  \coordinate (xm) at (2.5,0);
  \coordinate (xmn) at (2.5,2.1);
  \coordinate (xn) at (0,2.1);
  \draw[thick,->] (x) -- (xm);
  \draw[thick,->] (xm) -- (xmn);
  \draw[thick,->] (xmn) -- (xn);
  \draw[thick,->] (xn) -- (x);
  \fill (x) circle (1.5pt);
  \fill (xm) circle (1.5pt);
  \fill (xmn) circle (1.5pt);
  \fill (xn) circle (1.5pt);
  \node[below left] at (x) {$x$};
  \node[below right] at (xm) {$x+a\hat\mu$};
  \node[above right] at (xmn) {$x+a\hat\mu+a\hat\nu$};
  \node[above left] at (xn) {$x+a\hat\nu$};
  \node[below] at (1.25,0) {$U_{x\mu}$};
  \node[right] at (2.5,1.05) {$U_{x+a\hat\mu,\nu}$};
  \node[above] at (1.25,2.1) {$U^\dagger_{x+a\hat\nu,\mu}$};
  \node[left] at (0,1.05) {$U^\dagger_{x\nu}$};
\end{tikzpicture}
\caption{One oriented Wilson plaquette. The arrows follow the ordered
product in Eq.~\eqref{eq:su2-plaquette}; daggered links traverse opposite
to their positive coordinate directions. Its contribution to the action is
weighted by the corresponding term in Eq.~\eqref{eq:wilson-plaquette-action}.}
\label{fig:wilson-plaquette}
\end{figure}

The nonlinear Wilson-gauge vertices cited below use the bare-coupling
convention $g_0$ and generators $T^a=\sigma^a/2$. The link and oriented
plaquette variables are
\begin{align}
 U_{x\mu}&=\exp[-iag_0 A^a_{x\mu}T^a],
 \label{eq:su2-link}\\
 U_{x,\mu\nu}&=U_{x\mu}U_{x+a\hat\mu,\nu}
 U^\dagger_{x+a\hat\nu,\mu}U^\dagger_{x\nu}.
 \label{eq:su2-plaquette}
\end{align}
The Wilson action is
\begin{equation}
 S_g=\frac{4}{g_0^2}\sum_x\sum_{\mu<\nu}
 \left(1-\frac12\operatorname{Re}\Tr U_{x,\mu\nu}\right),
 \label{eq:wilson-plaquette-action}
\end{equation}
which has continuum normalization $\frac14\int\dd^4x\,
F^a_{\mu\nu}F^a_{\mu\nu}$ for this generator convention
\cite{Wilson1974}. Figure~\ref{fig:wilson-plaquette} depicts the ordered
link product. This action specifies the perturbative vertex-generator
convention; it does not imply that a full gauge-sector determinant or
physical gauged decay prefactor has been computed.

\subsection{Contained-background gauge--Goldstone diagnostic}
\label{subsec:su2-coupled-diagnostic}

A separate diagnostic run dated September 30, 2026, tested the coupled
gauge--Goldstone and ghost operators on a contained Gaussian profile. This
profile is not a stationary Coleman bounce, and the run is distinct from the
$L=26$ on-shell Ward-identity test above; its spectrum is an operator check,
not a determinant ratio or a decay-rate calculation.

In background 't Hooft--Feynman gauge, $\xi=1$, the matrix-free block for
each of the three identical $SU(2)$ generators acts on
$(A_1,A_2,A_3,A_4,\chi)$:
\begin{equation}
\mathcal H_{A\chi}=
\begin{pmatrix}
(-\Delta+m_W^2)\delta_{\mu\nu} & 2q\,\partial_\mu\rho\\
2q\,\partial_\nu\rho & -\Delta+V_T''+m_W^2
\end{pmatrix}.
\label{eq:su2-coupled-hessian}
\end{equation}
Here $q=g/2$, $m_W^2=q^2\rho^2$, $\mu,\nu=1,\ldots,4$, and $V_T''$ denotes the transverse scalar
curvature in the coupled block. Together with the radial Higgs block, the
three five-component sectors give a bosonic Hessian of dimension
$16L^4=65{,}536$ at $L=8$. The Faddeev--Popov ghost operator is evaluated
separately because it enters the functional determinant with a different
power.

The run used $a=2.0\times10^{-3}\,\mathrm{GeV}^{-1}$ and $g=0.65$.
Table~\ref{tab:su2-diagnostic} summarizes its setup and numerical check.
The displayed gauge and ghost spectra are the 16-mode windows for one
representative generator; the radial background makes the three gauge blocks
identical.

\begin{table}[t]
\caption{Settings and checks for the September 30, 2026, contained-background
$SU(2)$ diagnostic. This run is separate from the scalar production lattice.}
\label{tab:su2-diagnostic}
\begin{ruledtabular}
\footnotesize
\begin{tabular}{@{}p{0.34\columnwidth}p{0.62\columnwidth}@{}}
Quantity & Setting or result \\
\hline
Lattice & $L=8$, $a=2.0\times10^{-3}\,\mathrm{GeV}^{-1}$; 4096 sites \\
Background & Contained Gaussian; reported boundary fraction $1.495\times10^{-5}$ \\
Gauge & $g=0.65$, $\xi=1$, hence $q=0.325$ \\
Bosonic Hessian & Higgs plus three $(A_\mu,\chi)$ blocks; $65{,}536\times65{,}536$ \\
Spectrum & 16 gauge--Goldstone and 16 ghost modes for one generator \\
Self-adjointness & Bilinear-form residual $2.26\times10^{-16}$ \\
\end{tabular}
\end{ruledtabular}
\end{table}

\begin{figure*}[t]
\centering
\begin{minipage}[t]{0.485\textwidth}
\centering
\begin{tikzpicture}
\begin{axis}[
  width=0.95\linewidth,height=0.25\textheight,
  xlabel={Mode index},ylabel={Eigenvalue ($\mathrm{GeV}^2$)},
  title={Coupled gauge--Goldstone block},grid=major,
  legend pos=north west,scaled y ticks=base 10:3]
\addplot+[mark=o,thick] coordinates {
(1,20.071463)(2,20.071463)(3,20.071463)(4,20.071463)
(5,24642.5598)(6,145805.706)(7,146352.938)(8,146352.938)
(9,146422.659)(10,146422.659)(11,146444.091)(12,146444.091)
(13,146448.278)(14,146448.278)(15,146466.994)(16,146466.994)};
\addlegendentry{Gaussian background}
\addplot+[mark=square,thick] coordinates {
(1,0)(2,0)(3,0)(4,0)(5,25000)(6,146446.609)
(7,146446.609)(8,146446.609)(9,146446.609)(10,146446.609)
(11,171446.609)(12,171446.609)(13,171446.609)(14,171446.609)
(15,171446.609)(16,292893.219)};
\addlegendentry{False vacuum}
\end{axis}
\end{tikzpicture}
\end{minipage}\hfill
\begin{minipage}[t]{0.485\textwidth}
\centering
\begin{tikzpicture}
\begin{axis}[
  width=0.95\linewidth,height=0.25\textheight,
  xlabel={Mode index},ylabel={Eigenvalue ($\mathrm{GeV}^2$)},
  title={Faddeev--Popov ghost operator},grid=major,
  legend pos=north west,scaled y ticks=base 10:3]
\addplot+[mark=o,thick] coordinates {
(1,61.194087)(2,146449.483)(3,146449.483)(4,146449.483)
(5,146467.795)(6,146467.795)(7,146467.795)(8,146467.795)
(9,146841.834)(10,292893.352)(11,292893.352)(12,292894.203)
(13,292894.203)(14,292900.48)(15,292900.48)(16,292902.607)};
\addlegendentry{Gaussian background}
\addplot+[mark=square,thick] coordinates {
(1,0)(2,146446.609)(3,146446.609)(4,146446.609)
(5,146446.609)(6,146446.609)(7,146446.609)(8,146446.609)
(9,292893.219)(10,292893.219)(11,292893.219)(12,292893.219)
(13,292893.219)(14,292893.219)(15,439339.828)(16,439339.828)};
\addlegendentry{False vacuum}
\end{axis}
\end{tikzpicture}
\end{minipage}
\caption{Low 16 eigenvalues of the coupled gauge--Goldstone block and the
Faddeev--Popov ghost operator for one generator in the contained $L=8$
Gaussian diagnostic. The circles denote the Gaussian-background spectra and
the squares the false-vacuum spectra; eigenvalues are in $\mathrm{GeV}^2$.
The three gauge--Goldstone blocks are identical. Since the profile is not a
stationary bounce, these spectra validate the operator implementation but do
not define a gauged decay prefactor.}
\label{fig:su2-coupled-spectra}
\end{figure*}
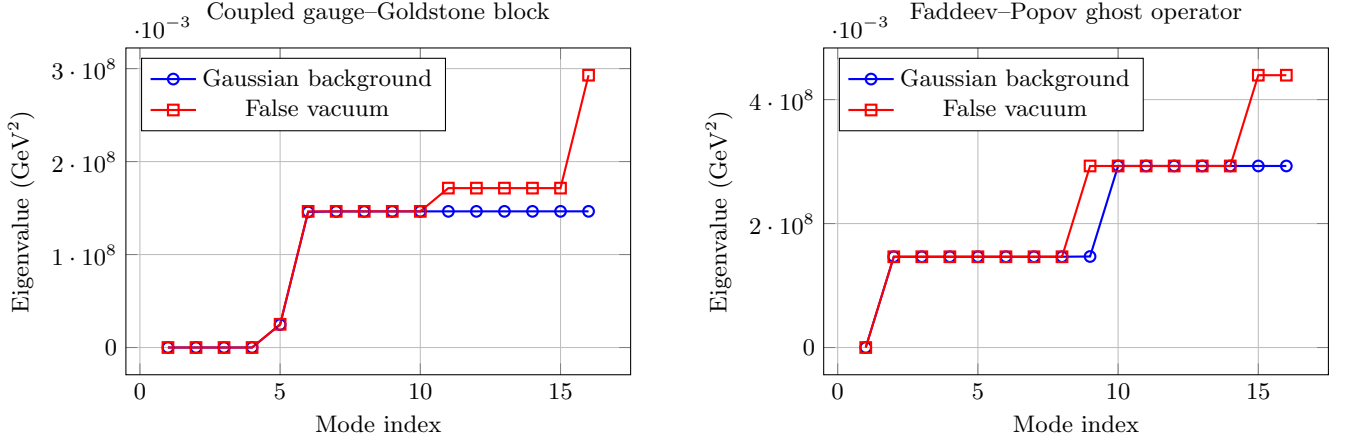

Figure~\ref{fig:su2-coupled-spectra} compares the low spectra with the
false-vacuum levels. The first four gauge--Goldstone eigenvalues on the
Gaussian background are $20.0715\,\mathrm{GeV}^2$, compared with four
near-zero false-vacuum gauge modes; the fifth is $24{,}642.6\,\mathrm{GeV}^2$
versus $25{,}000.0\,\mathrm{GeV}^2$. The lowest ghost eigenvalue is
$61.1941\,\mathrm{GeV}^2$, versus a near-zero false-vacuum mode. The four
near-zero false-vacuum gauge modes are constant periodic gauge
configurations, so a gauge-volume prescription is required before any
determinant ratio can be assigned a physical interpretation.

The local counterterm basis and continuum pole residues have been derived
and checked for algebraic closure; the reported scalar and background
Feynman-gauge pole checks close to $3.11\times10^{-15}$. Nonlinear scalar-link,
Wilson-gauge, and ghost vertex tables have been generated and independently
audited against finite transformations and separate action expansions at
tested routings. These results validate components of the implementation,
not a complete lattice-to-continuum matching calculation.

The latest matching work contains a provisional finite Higgs kinetic-slope
estimate, $\delta Z_H=0.0329117974\pm3.60\times10^{-7}$, where the quoted
uncertainty is QMC statistical error only. A gauge-linked six-point
insertion, $\delta\gamma_{\rm gauge}=-0.01887194\pm0.00058491$, and selected
higher-point projections are also pilots, not production-precision
coefficients. The continuum gauge/ghost two-point tensor, complete finite
gauge-sector coefficients, adequate regulator-cancellation and
Ward/Slavnov--Taylor closure, and the gauge-volume/collective-coordinate
normalization remain unfinished. No physical $SU(2)$ determinant or gauged
decay prefactor is reported.

Separate October 14--18 finite-lattice audits addressed translation-mode
bookkeeping and trace-estimator stability. A verified-subspace analysis found
four-translation tangent capture fractions of $90.8762\%$, $94.3902\%$,
and $96.1843\%$ in the $L=24,30,36$ spacing samples, respectively. In a
distinct paired-probe mode-exchange study, the 256-probe shifts at those
sizes agreed with the finite-mode bookkeeping predictions within $0.44$
paired standard errors. These analyses use different profiles and probe
streams and are not a common spacing extrapolation; they check finite-mode
bookkeeping rather than the full trace--log or a continuum limit.

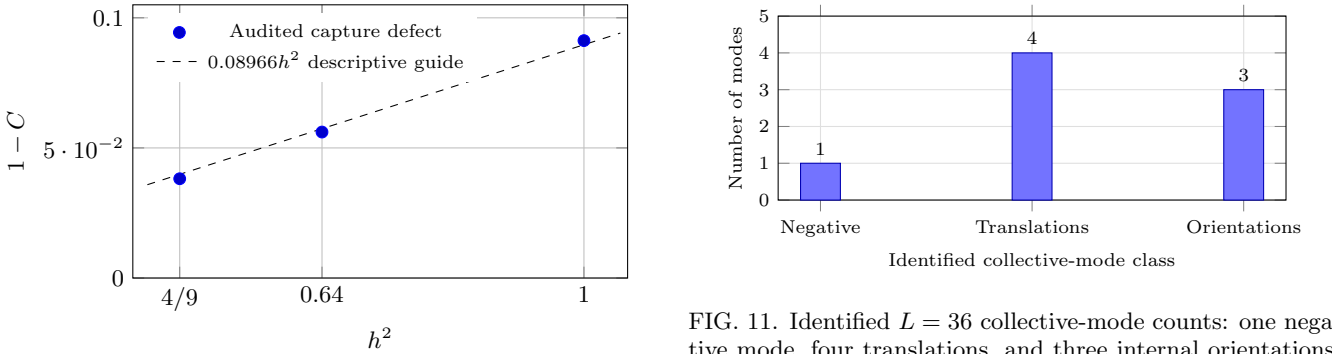
\begin{figure}[t]
\centering
\begin{tikzpicture}
\begin{axis}[
  width=0.94\columnwidth,height=0.22\textheight,
  xlabel={$h^2$},ylabel={$1-C$},
  xmin=0.38,xmax=1.06,ymin=0,ymax=0.105,
  xtick={0.44444,0.64,1.0},
  xticklabels={$4/9$,$0.64$,$1$},
  grid=major,
  legend style={font=\scriptsize,draw=none,at={(0.03,0.97)},anchor=north west}]
\addplot+[only marks,mark=*,mark size=2.2pt] coordinates {
  (0.444444,0.038157) (0.64,0.056098) (1.0,0.091238)};
\addlegendentry{Audited capture defect}
\addplot+[black,dashed,no marks] coordinates {
  (0.4,0.035864) (1.05,0.094143)};
\addlegendentry{$0.08966h^2$ descriptive guide}
\end{axis}
\end{tikzpicture}
\caption{Four-translation tangent-capture defect from the verified-subspace
spacing audit at fixed $Lh=24$. The dashed line is the through-origin
descriptive guide $1-C=0.08966h^2$ from the three points; it has no
inferential uncertainty and is not a controlled continuum extrapolation.}
\label{fig:spacing-capture}
\end{figure}

Table~\ref{tab:spacing-subspace} gives the verified-subspace quantities at
fixed box extent $Lh=24$ in the production scaled units. The mode-exchange prediction is
$\Delta_{\rm ex}=\log[\lambda_B/|\lambda_T|]$, where $\lambda_T<0$ is the
fourth translation mode and $\lambda_B>0$ is the position-selected boundary
mode. Figure~\ref{fig:spacing-capture} shows the capture defect against
$h^2$; its dashed guide is descriptive only and is not a continuum
extrapolation.

\begin{table*}[t]
\caption{Verified-subspace spacing diagnostics for the fixed box extent $Lh=24$ in the production scaled units. $C$ is the four-translation tangent capture fraction,
$\lambda_T$ is the fourth translation eigenvalue, $\lambda_B$ is the
position-selected positive boundary eigenvalue, and $\Delta_{\rm ex}$ is
the predicted finite-mode exchange. These data audit mode bookkeeping and
do not constitute a full-determinant spacing extrapolation.}
\label{tab:spacing-subspace}
\begin{ruledtabular}
\footnotesize
\begin{tabular}{@{}rrrrrrr@{}}
$L$ & $h$ & $C$ ($\%$) & $(1-C)/h^2$ & $\lambda_T$
& $\lambda_B$ & $\Delta_{\rm ex}$ \\
\hline
24 & $1$ & 90.8762 & 0.09124 & $-3.23612\times10^{-3}$ & 0.268131 & 4.41710 \\
30 & $0.8$ & 94.3902 & 0.08765 & $-2.12606\times10^{-3}$ & 0.273198 & 4.85592 \\
36 & $2/3$ & 96.1843 & 0.08585 & $-2.46598\times10^{-4}$ & 0.281746 & 7.04100 \\
\end{tabular}
\end{ruledtabular}
\end{table*}

\subsection{\texorpdfstring{$L=36$}{L=36} field and collective-mode diagnostics}
\label{subsec:l36-diagnostics}

\setlength{\textfloatsep}{8pt}
\setlength{\floatsep}{5pt}

The separate $L=36$ run report provides a central line profile and an
$x$--$y$ slice through the relaxed field, with the other coordinates fixed
at the lattice midpoint. These are finite-lattice spatial diagnostics and
are not merged with the $L=26$ production determinant results. A separate
$L=36$ low-mode audit recovered the four-dimensional longitudinal
translation cluster and one negative longitudinal mode. The transverse
orientation mode is counted once in each of the three degenerate transverse
channels, giving three internal-orientation modes in the full field. The
bar chart in Fig.~\ref{fig:l36-mode-counts} counts only these identified
collective modes, not the full spectrum.

\begin{figure}[t]
\centering
\begin{tikzpicture}
\begin{axis}[
  ybar,bar width=15pt,
  width=0.96\columnwidth,height=0.17\textheight,
  ymin=0,ymax=5,
  xlabel={Identified collective-mode class},
  ylabel={Number of modes},
  xtick={1,2,3},
  xticklabels={Negative,Translations,Orientations},
  xticklabel style={font=\scriptsize,align=center},
  ytick={0,1,2,3,4,5},
  nodes near coords,
  nodes near coords style={font=\scriptsize},
  grid=major,grid style={gray!25},
  tick label style={font=\scriptsize},
  label style={font=\scriptsize}]
\addplot[fill=blue!55,draw=blue!70!black] coordinates
  {(1,1)(2,4)(3,3)};
\end{axis}
\end{tikzpicture}
\caption{Identified $L=36$ collective-mode counts: one negative mode, four
translations, and three internal orientations. Other eigenpairs are omitted.}
\label{fig:l36-mode-counts}
\end{figure}

\section{Reproducibility and software}
\label{sec:reproducibility}

The principal scalar implementation is the single-file program
\texttt{Higgs\_fluctuation\_solver\_stabilized.py}; its nonlinear vertex
library is \texttt{su2\_lattice\_vertices.py}. The program separates
deterministic analytic validation from the more expensive research and
production stages. Its validation mode compares a matrix-free periodic
operator against known Fourier levels and checks complete degeneracies before
the bounce calculation. The production scalar stages can be run separately
or sequentially, with output summaries and restartable checkpoints. The
numerical values in this manuscript are transcribed from the project guide
and associated run reports; exact source revision, checkpoint, probe streams,
and machine-readable summaries to be archived together during
submission. The separate $L=8$ contained-background gauge--Goldstone and
ghost diagnostic is documented in \texttt{SU2\_Gauge\_Coupled\_Report.tex};
it is not part of the scalar production or spacing data sets.

\section{Discussion and conclusions}
\label{sec:conclusion}

The principal result of this work is a documented numerical chain from a
metastable four-component scalar potential through a continuum bounce,
finite-lattice stationary background, low-mode classification, stochastic
bare trace--log, and continuum-subtracted scalar determinant. Matrix-free
operator application makes the large lattice Hessians tractable, while
separate channel solves and explicit low-mode deflation preserve the
collective-coordinate structure needed by the determinant calculation.

The benchmark does not yet establish a controlled continuum or volume limit.
The bare SLQ result is regulator dependent, and the quoted uncertainty on the
continuum-subtracted trace--log is not a fully calibrated error estimate.
More fundamentally, the $\rho^8$ leading potential is not closed under its
scalar one-loop divergence: the local basis must include $\rho^{10}$ and
$\rho^{12}$ operators, whose finite coefficients remain unmatched. The
reported continuum determinant is therefore an intermediate
subtraction-prescription result rather than a complete matched one-loop EFT
prediction. The scalar partial prefactor also retains a missing
internal-orientation normalization and requires a physical units convention
before it can be interpreted as a rate. These boundaries are part of the
result: they separate what the present pipeline computes from what remains a
research task.

For the gauge extension, the implemented operators, Ward checks, pole
relations, and vertex audits provide prerequisites for a future calculation.
The separate $L=8$ contained-Gaussian test adds low-spectrum checks for the
coupled gauge--Goldstone and ghost operators, with a bilinear-form
self-adjointness residual of $2.26\times10^{-16}$. Because its profile is not
stationary, this is an operator-validation result, not a decay calculation.
They do not replace finite matching, a completed coupled gauge--Goldstone--
ghost determinant, or tests of regulator, gauge-parameter, spacing, and
volume independence. Any later comprehensive study should report those
quantities only after the corresponding matching coefficients and determinant
have been completed and independently validated.

\clearpage
\begin{widetext}
\centering
\begin{minipage}{0.52\textwidth}
  \centering
  \includegraphics[width=\linewidth,height=0.22\textheight,keepaspectratio]
    {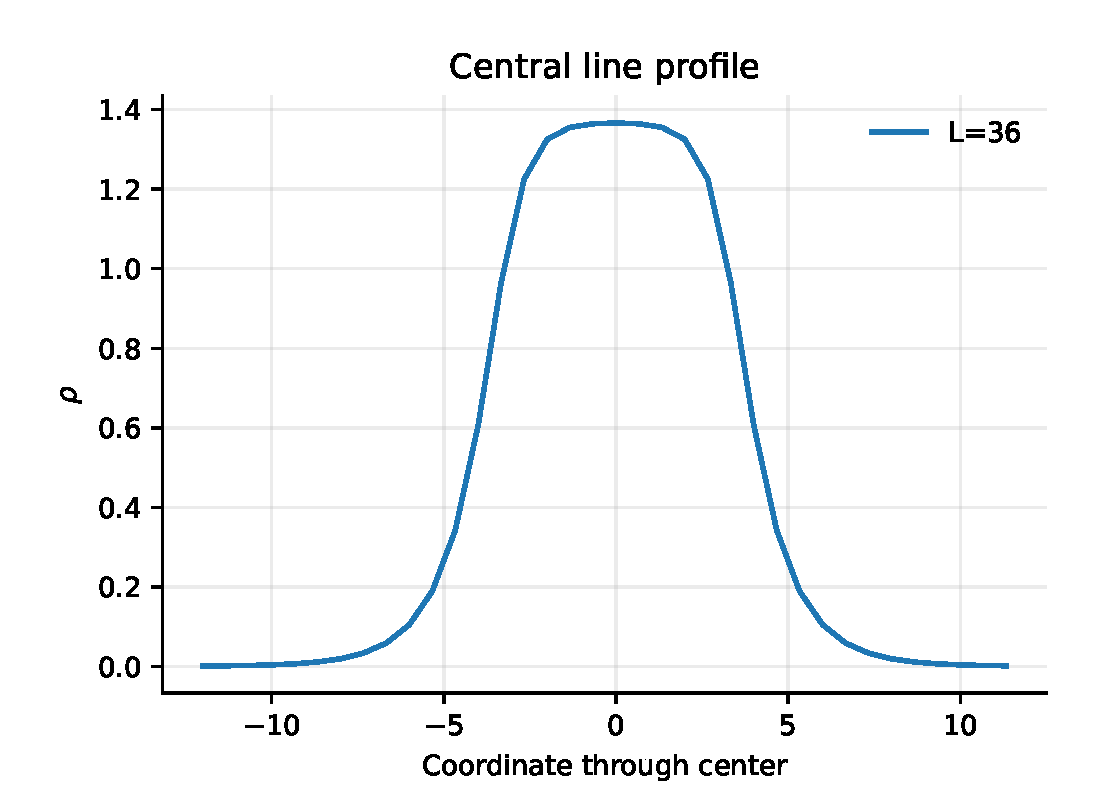}
\end{minipage}\hfill
\begin{minipage}{0.43\textwidth}
  \centering
  \includegraphics[width=\linewidth,height=0.22\textheight,keepaspectratio]
    {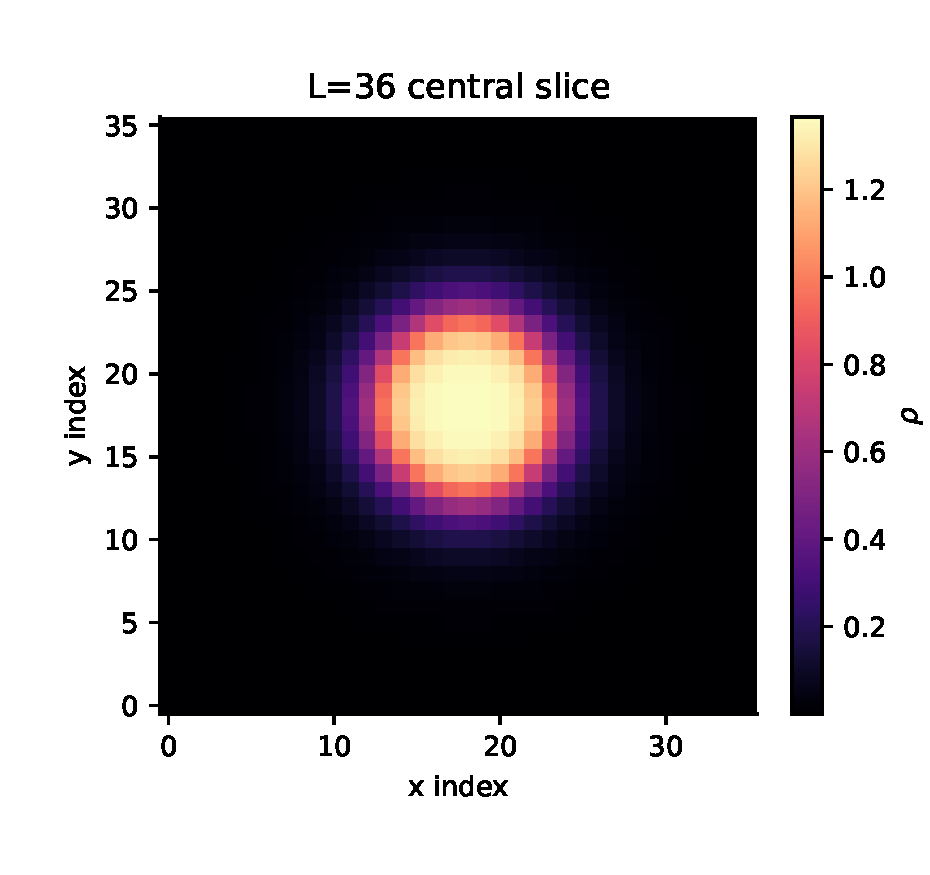}
\end{minipage}
\par\vspace{3pt}
\refstepcounter{figure}
\label{fig:l36-field-diagnostics}
\makeatletter
\@makecaption{\fnum@figure}{$L=36$ relaxed-field diagnostics: central line
profile (left) and central $x$--$y$ slice (right). These show finite-lattice
geometry, not a continuum extrapolation or an additional determinant result.}
\makeatother
\end{widetext}

\begin{acknowledgments}
OpenAI Codex provided organizational
and writing assistance. The author is responsible for verifying the
manuscript's technical content; the tool did not run the reported simulations.
\end{acknowledgments}

\noindent\textit{Data availability.}
The source and outputs are not yet archived; during submission, we will
deposit the code, dependencies, run metadata, probes, and result tables and
add the repository identifier.

\FloatBarrier
\begingroup
\renewcommand{\baselinestretch}{1.0}
\renewcommand{\bibfont}{\fontsize{8.5}{9.5}\selectfont}
\setlength{\bibsep}{0pt}

\endgroup

\end{document}